\documentclass[sigconf]{acmart} 

\AtBeginDocument{%
  \providecommand\BibTeX{{%
    \normalfont B\kern-0.5em{\scshape i\kern-0.25em b}\kern-0.8em\TeX}}}

\copyrightyear{2021}
\acmYear{2021}
\setcopyright{acmcopyright}\acmConference[WiPSCE '21]{The 16th Workshop in Primary and Secondary Computing Education}{October 18--20, 2021}{Virtual Event, Germany}
\acmBooktitle{The 16th Workshop in Primary and Secondary Computing Education (WiPSCE '21), October 18--20, 2021, Virtual Event, Germany}
\acmPrice{15.00}
\acmDOI{10.1145/3481312.3481345}
\acmISBN{978-1-4503-8571-8/21/10}

\usepackage{xspace}
\usepackage{tabularx}
\usepackage{colortbl}
\usepackage{subfig}
\usepackage{multirow}
\usepackage{booktabs}
\usepackage{listings}
\usepackage{scratch3}

\setscratch{scale=.60}

\newcommand\definetool[2]{\newcommand{#1}{{\textsc{#2}}\xspace}}
\definetool{\Scratch}{Scratch}
\definetool{\leila}{LeILa}
\definetool{\whisker}{Whisker}
\definetool{\litterbox}{LitterBox}
\definetool{\bastet}{Bastet}
\definetool{\scratchblocks}{scratchblocks}

\colorlet{punct}{red!60!black}
\definecolor{background}{HTML}{EEEEEE}
\definecolor{delim}{RGB}{20,105,176}
\colorlet{numb}{magenta!60!black}

\lstdefinelanguage{json}{
    basicstyle=\normalfont\ttfamily,
    numbers=left,
    numberstyle=\scriptsize,
    stepnumber=1,
    numbersep=8pt,
    showstringspaces=false,
    breaklines=true,
    frame=lines,
    backgroundcolor=\color{background},
    literate=
     *{0}{{{\color{numb}0}}}{1}
      {1}{{{\color{numb}1}}}{1}
      {2}{{{\color{numb}2}}}{1}
      {3}{{{\color{numb}3}}}{1}
      {4}{{{\color{numb}4}}}{1}
      {5}{{{\color{numb}5}}}{1}
      {6}{{{\color{numb}6}}}{1}
      {7}{{{\color{numb}7}}}{1}
      {8}{{{\color{numb}8}}}{1}
      {9}{{{\color{numb}9}}}{1}
      {:}{{{\color{punct}{:}}}}{1}
      {,}{{{\color{punct}{,}}}}{1}
      {\{}{{{\color{delim}{\{}}}}{1}
      {\}}{{{\color{delim}{\}}}}}{1}
      {[}{{{\color{delim}{[}}}}{1}
      {]}{{{\color{delim}{]}}}}{1},
}

\usepackage{fbox}
\newcommand{\summary}[2]{
        \vspace{2mm}
        \noindent
        \fbox{%
            \parbox{.97\linewidth}{%
                    \textbf{#1 Summary.}
                #2
            }%
        }%
}%

\usepackage[colorinlistoftodos,prependcaption,textsize=tiny]{todonotes}

\begin{document}

\title{Data-driven Analysis of Gender Differences and Similarities\\ in \Scratch Programs}

\author{Isabella Graßl}
\email{isabella.grassl@uni-passau.de}
\orcid{1234-5678-9012}
\affiliation{%
  \institution{University of Passau}
  \state{Passau}
  \country{Germany}
}

\author{Katharina Geldreich}
\email{katharina.geldreich@tum.de}
\affiliation{%
  \institution{Technical University of Munich}
  \city{Munich}
  \country{Germany}}

\author{Gordon Fraser}
\email{gordon.fraser@uni-passau.de}
\affiliation{%
  \institution{University of Passau}
  \city{Passau}
  \country{Germany}
}

\renewcommand{\shortauthors}{Gra{\ss}l, et al.}

\begin{abstract}
	%
%
Block-based programming environments such as \Scratch are an essential entry point to computer science. In order to create an effective learning environment that has the potential to address the gender imbalance in computer science, it is essential to better understand gender-specific differences in how children use such programming environments. 
%
In this paper, we explore gender differences and similarities in \Scratch programs along two dimensions:
In order to understand what \emph{motivates} girls and boys to use \Scratch, we apply a topic analysis using unsupervised machine learning for the first time on \Scratch programs, using a dataset of 317 programs created by girls and boys in the range of 8--10 years. In order to understand how they \emph{program} for these topics, we apply automated program analysis on the code implemented in these projects.
%
We find that, in-line with common stereotypes, girls prefer topics that revolve around unicorns, celebrating, dancing and music, while boys tend to prefer gloomy topics with bats and ghouls, or competitive ones such as soccer or basketball. 
%
Girls prefer animations and stories, resulting in simpler control structures, while boys create games with more loops and conditional statements, resulting in more complex programs. Considering these differences can help to improve the learning outcomes and the resulting computing-related self-concepts, which are prerequisites for developing a longer-term interest in computer science.
%
%
%
%
\end{abstract}

\ccsdesc[500]{Social and professional topics~Gender}
\ccsdesc[500]{Software and its engineering~Software development techniques}
\ccsdesc[500]{Social and professional topics~Computing education}

\keywords{Scratch, gender, topic modeling, automated code analysis.}
\maketitle

\section{Introduction}
\label{sec:intro}

Even though the promotion of girls in computer science (CS) is increasingly emphasized, girls remain underrepresented~\cite{murphy2019, wang2019}. 
A potential approach to address this problem is to provide girls insights into CS~\cite{sharma2020improving} and thus to arouse their interest in it before gender-specific role attributions become entrenched during puberty~\cite{beyer2003, master2020cultural}.
In particular, the potential expression of creativity enabled by programming environments such as \Scratch \cite{resnick2009} is assumed to help exciting girls about programming~\cite{hubwieser2016, roque2016}. 
However, the continuing gender imbalance in CS 
suggests a need to better understand how girls program in \Scratch~\cite{hsu2014, aivaloglou2016, funke2017a, aivaloglou2019a}, in order to improve learning environments~\cite{aivaloglou2016, hubwieser2016} to consider gender-specific interests while ensuring the same learning outcomes. 

In this paper, we provide further evidence for gender differences and similarities in \Scratch by analyzing programs created by girls and boys in the range of 8--10 years.\footnote{We explicitly do not support binary gender thinking, however, the children categorized themselves as female and male.} The dataset contains 127 projects originating from a prior study by Funke and Geldreich~\cite{funke2017a}, in which they manually evaluated programs. We replicated~\cite{shepperd2018role} the programming course to obtain 192 additional projects; in total this combined dataset contains 319 \Scratch programs of which we used 317 for an automated analysis, created by 64 girls and 68 boys.

To encourage children's interest in learning to program, it is essential to motivate them for this new subject in the first place. The learning process is particularly sustainable when the children's interests are addressed and they also have fun \cite{aivaloglou2016, roque2016}. Especially when it comes to attracting girls, it is purposeful to make them aware of programming through their thematic preferences \cite{kelleher2007, hubwieser2016, sullivan2016}. Therefore, our first research question is as follows: 

\smallskip
\noindent\textbf{RQ1: } \textit{What gender differences and similarities can be identified in the topics chosen by girls and boys?}
\smallskip

\noindent We use the unsupervised machine learning method of Latent Dirichlet Allocation (LDA)~\cite{blei2003} to automatically extract topics in \Scratch programs in order to determine the thematic preferences of girls and boys. 
These extracted topics help to see what the children interest is and thus to design programming lessons that are equally attractive to girls and boys, or specifically attractive to girls.

While initial enthusiasm for programming is essential, learning outcomes are equally important as they determine the resulting computing-related self-concept and perceived self-efficacy~\cite{shavelson1976, vrieler2020computer}. 
These are prerequisites for a longer-term interest in CS and whether learners consider themselves suitable for studies in CS. In order to evaluate the learning outcomes, we consider the source code:

\smallskip
\noindent\textbf{RQ2: } \textit{What gender differences and similarities can be identified in the implementations of the programs?}
\smallskip

\noindent We use automated code analysis 
to provide insights into four dimensions of code: block types, programming concepts, code complexity, and code smells.
The identification of the preferred block types, i.e., the program type, and the associated programming concepts, is important for creating appealing learning scenarios and for making the learning needs of certain programming concepts apparent~\cite{adams2012, funke2017a, aivaloglou2019}. 
The presence of code smells has a negative impact on the performance of the children, the usability of the programs and the understanding of programming \cite{hermans2016, hermans2016a, fradrich2020, p.rose2020}.
Similarly, increased code smells and bugs are a source of frustration that can discourage children from continuing to program, and could also indicate inadequate teaching approaches.
Code complexity can be used as an indicator to see whether girls and boys are implementing correct programs differently, where individual learning is still needed and how their activities relate to learning objectives.

In detail, the contributions of this paper are as follows:
\begin{enumerate}
\item[1.] A replication of a \Scratch course~\cite{funke2017a} to obtain a combined dataset of 319 projects with gender labels (Section~\ref{sec:methodology}). 
\item[2.] An automated identification of topics in programs of girls and boys using unsupervised machine learning %
 (Section~\ref{sec:rq1}).
\item[3.] An automated code analysis of  \Scratch programs in relation to gender, hence the confirmation and extension of previous research results regarding the gender-specific use of code blocks and programming concepts (Section~\ref{sec:rq2}).
\item[4.] A discussion of didactic implications for designing universally suitable programming courses and tasks to promote girls in CS in particular (Section~\ref{sec:discussion}).
\end{enumerate}

Our analyses confirm stereotypical gender-differences regarding the choice of topics, and a tendency that girls produce programs that are more sequential and less complex. However, the differences in code are not a result of the chosen topics per se, but rather a result of the types of programs preferred: Girls prefer programming elements that serve to develop animations or stories, while boys mainly implement games in \Scratch. Consequently, programming courses need to be designed such that both topic preferences and application of programming concepts can be adequately addressed~\cite{dasgupta2018wide}.
%
%

\section{Background and Related Work}
\label{sec:background}

\Scratch~\cite{resnick2009} is one of the most popular introductory programming languages. 
To make it easier for programming beginners, 
 code is constructed with the help of blocks, which represent instructions of the program. Learning material often explicitly focuses more on creativity than on programming concepts~\cite{amanullah2019}. It has been shown that different types of projects result in the use of different programming concepts, e.g., animations represent a lower level of difficulty than projects implementing games~\cite{moreno-leon2020}.

Although recently research has started investigating gender differences in programming concepts in \Scratch~\cite{rubio2015a, fields2017, papavlasopoulou2020}, overall there has been relatively little explicit research on this topic, in particular in the context of \Scratch, which might be due to the fact that gender is only a self-reported information on the \Scratch website. Most publications manually analyzed programs from programming courses for their use of programming concepts and blocks. The consensus of the research is relatively clear: While girls prefer narrative structures and dialog messages in their programs, i.e., animations and stories, 
boys' programs are dominated by game-specific blocks and concepts such as \textit{motion} blocks, boolean expressions and conditional statements~\cite{adams2012, hsu2014, aivaloglou2016, fields2017, funke2017a}. 

This paper builds on a prior dataset of 127 \Scratch programs 
of Funke and Geldreich~\cite{funke2017a}, whose manually extracted results corroborate the previous results. 
 Girls are more likely to program stories using a large number of \textit{looks} blocks, which require a lower level of comprehension, and boys are more likely to program games, which require a large number of \textit{motion} blocks and have a higher level of comprehension. Overall, boys use almost twice as many different blocks as girls, which indicates that boys are more willing to experiment or that girls need fewer different blocks to accomplish their objectives.
Our work automates this acquisition process and extends it to analyze topics, code smells and complexity metrics, since these subjects in relation to gender have not yet been scientifically studied in an automated manner.
Gender differences have also been investigated in the domain of robot programming, where
 boys were shown to perform better than girls in tasks related to advanced programming concepts~\cite{sullivan2013, sullivan2016}. In addition, differences in their learning approach and methodology~\cite{rubio2015a, papavlasopoulou2020} were evident.

Overall, research in this direction helps to sustainably strengthen the interest of girls in CS. This is relevant because an increased proportion of women in CS has a positive effect on the productivity, communication and spirit of software teams \cite{vasilescu2015b, catolino2019, russo2020}. In addition, this mitigates the still prevailing gender bias in CS~\cite{wang2019a, wang2020b}.

\section{Method}

\label{sec:methodology}

\subsection{Data Collection}
To extend the results of Funke and Geldreich~\cite{funke2017a} and their original dataset, their introductory programming course was replicated by using the same materials and instructions.
The course was conducted with a total of nine groups of children with 8--10 children as an extracurricular offer at the Technical University of Munich. One group attended the course as a voluntary vacation activity, and the remaining eight were attended by four whole elementary school classes---one of 3rd grade and three of 4th grade---each of which was divided into two groups. Overall, 36 boys and 38 girls aged 8--10 years took part who had little to no experience in programming.

The programming course consists of three sessions (each for 3.5 hours) over three consecutive days and was conducted by a course instructor with a background in CS education. 
In the course, the children started by describing algorithms in natural language. At first they were introduced to \Scratch in an \textit{unplugged} way~\cite{bell2015cs} by allowing them to arrange haptic programming blocks corresponding to the programming commands in \Scratch into a script. Afterwards, the basic functions of \Scratch, such as repetitions and conditional statements, were introduced to the children in a learning circle based on the theme of a circus performance. In the process, they had to solve suitable tasks, which were, for example, to fill the circus ring with an audience, to have the ringmaster say a greeting, to have a tiger and a horse run back and forth or to have the tightrope walker balance. In the last unit of the course, the students were tasked to implement their own program and to present it. After filling in a planning sheet for their individual projects, they had three hours to realize their own ideas in \Scratch. The programs had to meet several requirements: They should contain more than one sprite, move at least one sprite during execution, and include at least one iteration as well as a conditional statement. Students were allowed to create multiple projects if they wanted to.  A detailed description of the development and progression of the course can be found in the description by Geldreich et al.~\cite{geldreich2019}.


\subsection{Dataset}
The dataset used here is the result of the last task of the course, where the students had to design and implement a \Scratch project based on a topic of their own choice. Since our data is based on an exact replication of the course by Funke and Geldreich~\cite{funke2017a} we combined their original data with the data we collected for our analysis. The original dataset consists of 127 projects from 26 girls and 32 boys, and we gathered another 192 projects from 38 girls and 36 boys. Accordingly, the dataset contains 319 projects---171 of them by girls and 148 by boys.
We removed projects that contain no modifications to the default project (i.e., the only content is the \Scratch cat as default sprite with no code) and thus obtain a dataset of 317 projects, 171 from 64 girls and 146 from 68 boys. Even programs without blocks but with, for instance, self-painted backgrounds provide information about how children use \Scratch and were therefore included in the analysis. The dataset contains only the code and information as given above (ID, age, gender) and no further annotation or demographic data.

\subsection{Data Analysis}

\subsubsection{RQ1}
We aim to automatically extract semantic topics from unstructured text data of the \Scratch programs in order to determine which topics are common in the programs of girls and boys.
To the best of our knowledge, an automatic identification of topics of \Scratch programs has not yet been conducted, which explicitly distinguishes this work from previous research~\cite{funke2017a}.
The text corpus of documents (here \Scratch programs), which contains different terms that form a vocabulary (e.g., names of sprites etc.), serves as input for the LDA model. The number of the hidden topics are represented by manually predetermined multinominal distributions. These distributions are drawn over all terms from Dirichlet distributions and then for each document (i.e., \Scratch program) a distribution of multinomial distributions is derived from such a Dirichlet distribution. Thus, different topics according to their grouped terms are automatically assigned to each document.

To create the text corpus we extracted tokens from the \Scratch programs.
The programs in the course were created locally using the \Scratch 2 desktop application and not on the \Scratch website, which means that the only available source of information is the code of the programs, but not other meta information such as instructions or comments. The \Scratch programs are in \texttt{.sb2} format, which is a zip archive format consisting of the media files used in the project together with the source code in JSON format. From these JSON files we extracted all tokens that represent names of sprites, costumes, backgrounds, sounds, variables, as well as all string inputs to common blocks (e.g., \textit{say} and \textit{think} blocks).

As usual when applying natural language processing (NLP), the textual data was preprocessed before serving as input for a machine learning model. 
For data cleaning, we normalized the text to lower case, and removed English and German stop words. In addition, we defined and removed customized stop words that represent the German and English default designations of the individual tokens, which do not add any content value for a topic identification: `stage', `bühnenbild', `pop', `plopp', `kostüm', `figur', `block', `hintergrund'. Furthermore, we removed digits and words with less than two characters as they do not contain relevant information.

One of the best established methods of unsupervised machine learning to generate unknown (``hidden'') topics from a large, unlabelled text dataset is Latent Dirichlet Allocation (LDA)~\cite{blei2003}, which is also increasingly  applied in software engineering~\cite{campbell2015latent}.
%
Based on common practice we used a minimum number of 10 occurrences for words when tokenizing and building a vocabulary, as words with fewer occurrences are likely not important enough in the data set.
Finally, the remaining tokens are used to create a sparse document-term-matrix of the programs.

The semantic analysis is implemented using the LDA model of the popular Scikit-Learn library\footnote{\url{https://scikit-learn.org/stable/modules/generated/sklearn.decomposition.LatentDirichletAllocation.html}}.
The LDA model is trained with the following parameter values, which are the commonly used default values: maximal learning iterations = 10, random state = 100, and a batch size of 128. 
A further parameter to choose is the number of topics, which influences the result quality of these topics. Since there is no standard way to choose the best model we empirically determined that 10 is a suitable number to cover different topic areas for a dataset of 317 projects; for larger numbers the topics were no longer semantically distinct and had a lot of overlap~\cite{jelodar2019latent}.

Using the model, each project is assigned a probability for each topic. The topic with the highest probability represents the \emph{dominant} topic of the project. We consider the dominant topics of the projects together with the gender information to analyze gender-specific differences in the topic distributions.

\subsubsection{RQ2}

At the level of code, we aim to investigate which programming blocks girls and boys prefer, which programming concepts they cover with these blocks, and how complex the resulting programs are. 
To investigate the distribution of covered programming statements and concepts, we extracted the opcodes of the blocks used directly from the JSON files.
The blocks are organized into categories such as \textit{looks} and \textit{motion} based on the ``drawers'' they are contained in within \Scratch. We compare projects in terms of the individual opcodes as well as these categories.
To sort blocks into the programming concepts they represent, we used the classification applied by Funke and Geldreich~\cite{funke2017a}. In contrast to their manual analysis, our procedure is automated and includes the additional metrics of code complexity and code smells. 

In order to measure the complexity of the programs we used \litterbox~\cite{fradrich2020}, which parses the JSON files into an abstract syntax tree, on which various metrics can be extracted. In particular we consider complexity in terms of the common Halstead and McCabe metrics. The Halstead metrics are based on the total and unique number of operators (e.g., keywords and tokens of the programming language) and operands (e.g., variables, literals, function names) in a program. The Halstead length represents the total number of operator occurrences and the total number of operand occurrences, the Halstead size (also known as vocabulary) is the total number of unique operators and operands. The volume represents the size (in bits) of space necessary for storing the program; the difficulty is proportional to the ratio of the total number of operands to the number of unique operands based on the intuition that if the same operands are used many times in a program, that program is more prone to errors. Finally, the Halstead effort measures the elementary mental discriminations. We implemented Halstead metrics for \litterbox while interpreting operators and operands as in previous work~\cite{moreno2016comparing}. The McCabe complexity (cyclomatic complexity) measures the number of linearly independent paths through a program's control flow graph. Since \Scratch programs often consist of many small, parallel scripts, we implemented an \emph{interprocedural} version of cyclomatic complexity (ICC) based on the interprocedural control-flow graph provided by \litterbox. Similar to lines of code (LOC) in textual programming languages we also measure size in terms of the number of blocks a program consists of. 

In order to detect code smells in the programs we used \litterbox. A code smell is a weakness in the code that requires refactoring as it might disturb the programming process~\cite{fowler1999}, which in turn results in frustration of the program, especially for beginners. In addition, the learning process is impaired, since the misconception is reinforced over time if no action is taken.
 
In contrast to the procedure for RQ1, no cleaning of names or any further cleaning steps were carried out in order not to falsify the results as information such as incremented sprite names (e.g., sprite1, sprite2) are relevant for the analysis of code smells.
 
%

\subsection{Threats to Validity}

\paragraph{Internal Validity}
Since there is neither an existing thematic framework nor labelled data, we had to apply unsupervised machine learning. In contrast to alternative approaches such as Latent Semantic Analysis~\cite{dumais1988} or word embeddings with Word2Vec~\cite{mikolov2013a} and subsequent clustering, LDA has the advantage of providing the probabilities of topic associations. Disadvantages of LDA are the necessity to decide on the number of topics in advance, lack of names of topics and reduced accuracy due to multilingual and short text. 
Since the programs are relatively small and created by novices,  complexity and code smells can only be seen as indicators. 
Other metrics such as the computational thinking score~\cite{moreno-leon2015} might have been considered, but there is no consensus about the most adequate metrics; furthermore these metrics are usually based on checking the existence of certain blocks in programs, which is conceptually similar to our measurement of programming concept occurrence.

\paragraph{External Validity}
The study does not claim to be exhaustive, as the 317 analyzed projects can only provide an insight. 
The study is limited by its participants due to their home resident as well as cultural and social background. 
Likewise, due to the format of the three-day workshop with the introductory tasks on the topic circus as well as the teachers present, there may have been a bias in the study conditions. In order to minimize this influence, the course was very carefully designed by didactics experts and has already been conducted several times. 

\paragraph{Construct Validity}
Further filtering may improve data quality, e.g., by excluding duplicates. For our ranking of topics, only the highest probability of a topic is considered, which might be misleading when the difference between two topics of a project is small. However, even setting a threshold would be subjective.

\section{Results}
\subsection{RQ1: Topic Analysis}
\label{sec:rq1}

\begin{table}[tb]
\small
\centering
\caption{The ten topics automatically generated from the LDA model with their most popular terms. Colors assigned to topics match Figures \ref{fig:lda2D} and
 \ref{fig:lda}.}
\label{tab:topicsCircus}
\vspace{-1em}
\begin{tabularx}{\linewidth}{p{0.05cm}X}
\toprule
 ID & Topic \\ \midrule
\cellcolor[HTML]{1f77b4} 0  & dog, bass, drum, elec, dance, snare, celebrate, right, bedroom, birthday               \\
\cellcolor[HTML]{ff7f0e} 1  &  meow, balloon, donut, unicorn, dance, rainbow, bear, magic, cave, party                 \\
\cellcolor[HTML]{2ca02c} 2  &  aufnahme, cymbal, drum, cave, spotlight, creak, alien, city, night, beat        \\
\cellcolor[HTML]{d62728} 3  & katze, klein, manege, castle, desert, bedroom, chalkboard, water                \\
\cellcolor[HTML]{9467bd} 4  & bat, ghoul, meow, stars, ghost, wizard, wall, brick, woods, neon            \\
\cellcolor[HTML]{8c564b} 5  &  miau, water, underwater, hallo, beach, hello, meow, city, malibu, puffs           \\
\cellcolor[HTML]{e377c2} 6  & cat, flying, soccer, ball, goal, dance, meow, head, nod, stars                               \\
\cellcolor[HTML]{7f7f7f} 7  &  ball, dee, adrian, seilt\"anzerin, bedroom, baseball, basketball, blue, sky, spotlight         \\
\cellcolor[HTML]{bcbd22} 8  & zirkusdirektor, m\"adchen, junge, affe, spotlight, zauberer, elefant, seilt\"anzerin, clown, seehund               \\
\cellcolor[HTML]{17becf} 9  & giga, pico, ghost, affe, drum, sky, blue, schon, hello, tom                       \\ \bottomrule
\end{tabularx}
\end{table}

\begin{figure}[tb]
	\centering
	\includegraphics[width=\linewidth]{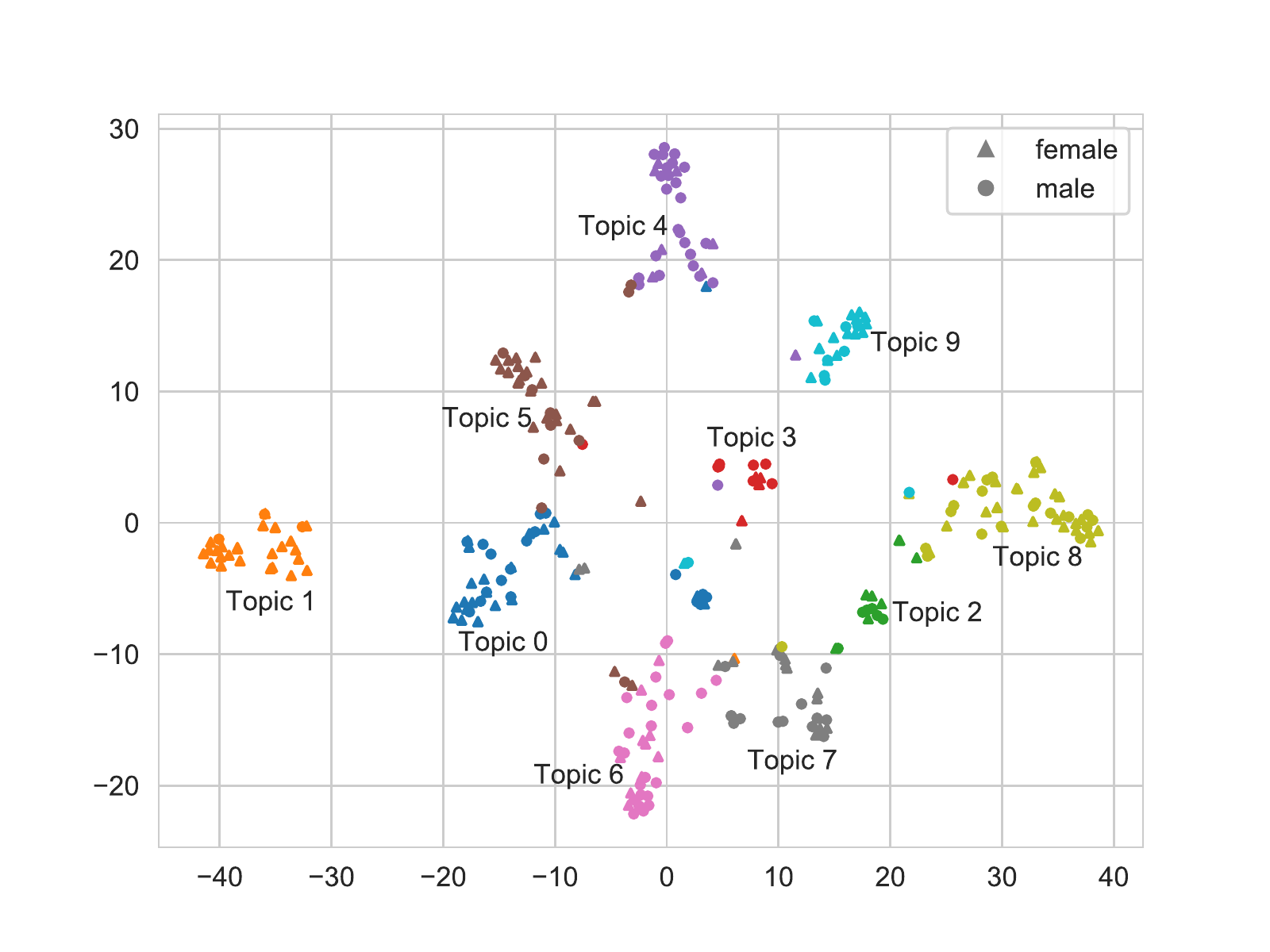}
	\vspace{-2em}
	\caption{The visualization of the ten topics generated from the LDA model. Each data point corresponds to a program.}
	\label{fig:lda2D}
\end{figure}

Table~\ref{tab:topicsCircus} lists the ten topics automatically generated from the LDA model with their most weighted terms of the \Scratch programs.
In order to observe the distribution of the topics more precisely, Figure~\ref{fig:lda2D} shows a visualization of these topics in the LDA model in a 2D space using the t-SNE (t-distributed stochastic neighbor embedding) algorithm~\cite{van2008}.
The arrangement of the topics in the space clearly shows to what extent they overlap and how similar they are to each other as the algorithm tries to minimize the Kullback-Leibler divergence of the between the joint probabilities of the data points. The algorithm preserves only nearest-neighbors and is therefore very sensitive to the choice of perplexity (here 15).

\begin{figure}[tb]
	\centering
	\includegraphics[width=\linewidth]{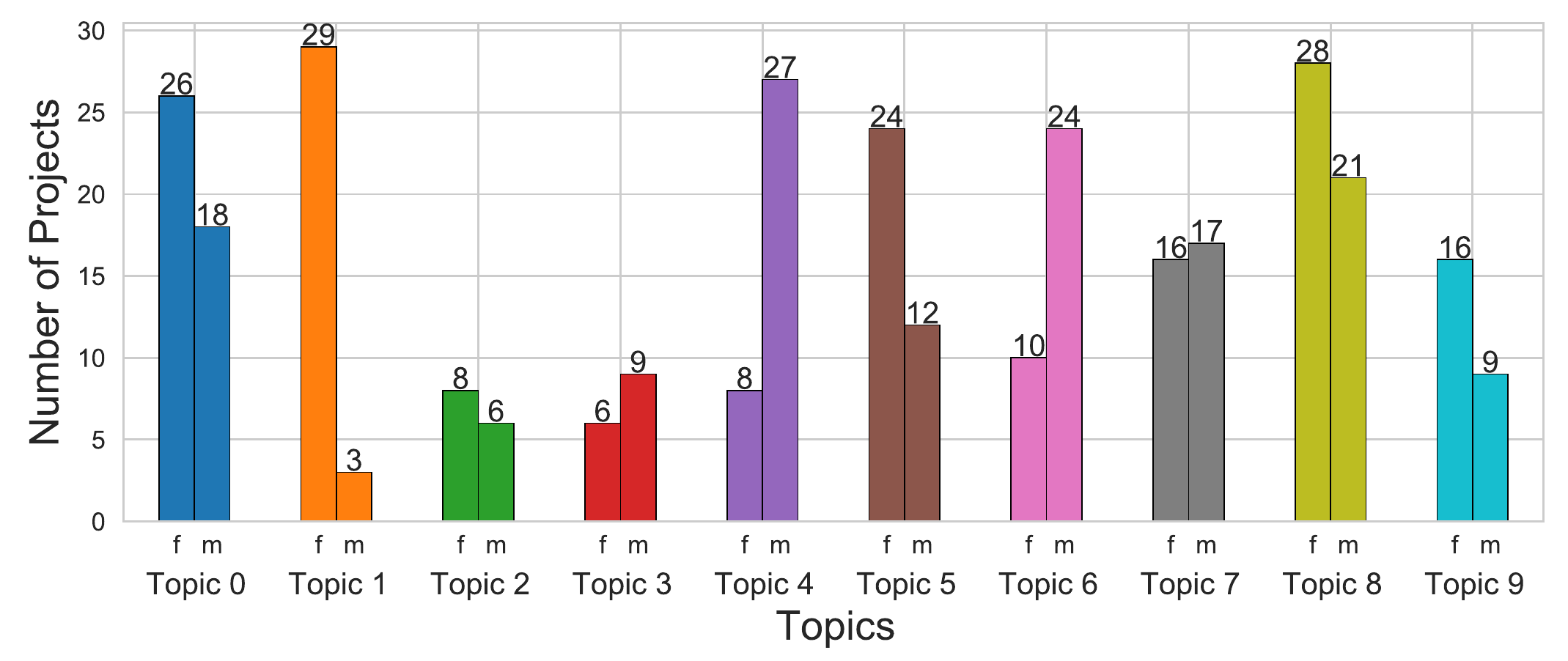}
	\vspace{-2em}
	\caption{The distribution of the automatically generated topics from the LDA model in girls' and boys' programs.}
	\label{fig:lda}
\end{figure}

Topics 1 and 4 seem to be independent and disjoint from all others, and the plot already hints at gender differences for these topics. The distribution of the derived topics among the genders is shown in detail in Figure~\ref{fig:lda}. 
For the other topics, Figure~\ref{fig:lda2D} shows moderate intra- and inter-cluster distances, which indicates a greater similarity or a lower delimitation of the topics. Topic~8, which represents the circus topic (i.e., the topic used in the preparatory exercises), shows a high dispersion, which means that it is a very wide field, while in Topics~4 or~6 the programs are very close together and therefore show a high inter-class similarity. Topic 3 has strong outliers which indicates that this topic is not highly homogeneous and addresses very different aspects of programs that are also reflected in other topics.
The visualization based on the LDA model provides indicative and apparent insights on how the programs are arranged in a high-dimensional space, hence a glance to what extent the model has identified latent patterns in the programs.

\subsubsection{Thematic Differences}

\begin{figure}[tb]
    \centering
    \subfloat[This girl's program (ID73) is 98 \% assigned to Topic 1.]{\includegraphics[scale=0.22]{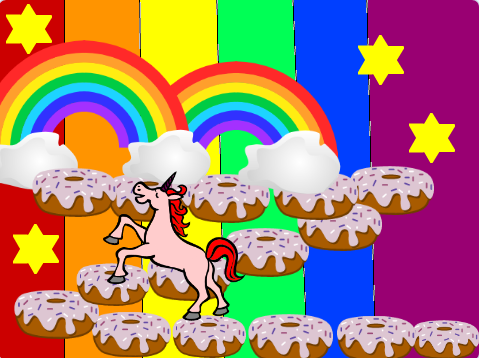}}
    \hfill
    \subfloat[This boy's program (ID55) is 97 \% assigned to Topic 4.]{\includegraphics[scale=0.22]{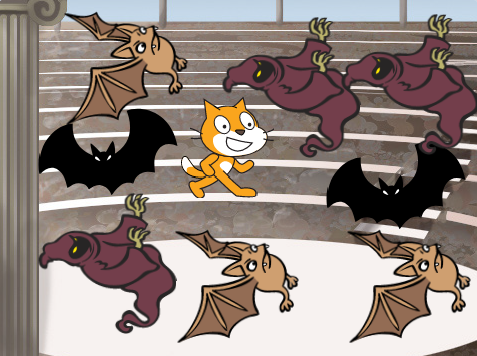}}
    \caption{Two representative program examples, which have the highest allocation of the LDA model of the top-topics.}%
    \label{fig:rq1topicexample}%
\end{figure}

The most popular topics in \Scratch programs differ between girls and boys (Fig. \ref{fig:lda}). 
In approximately 17 \% of all girls' projects, the semantic field of a \textit{fantasy world} is formed in Topic 1 as dominant topic by the terms \textit{unicorn}, \textit{rainbow} and \textit{magic} (Table~\ref{tab:topicsCircus}). 
These keywords fit into the social attribution of girls' fascination with unicorns, as especially in recent years as the enthusiasm for unicorns has increased in pop culture. 
Additionally, there seems to be cause for celebration, which the terms \textit{balloon}, \textit{donut}, \textit{dance} and \textit{celebrate} indicate. 
Figure \ref{fig:rq1topicexample} shows an example project (ID73, Unicorn loves Rainbow) which is assigned to Topic 1 by 98 \% and was created by a girl. In the program, a unicorn moves from the right edge of the screen on static donuts to dance music in zigzag movements to the left edge of the screen and back. Towards the end of the program it says ``Hello''. The sprites Donut and Cloud contain no code. The Rainbow and Star sprites both contain the script ``When flag clicked, repeat 100, change effect color''. 

Strikingly, Topic 1 is the least popular topic for boys (approx. 2~\%). Instead, the terms \textit{bat}, \textit{ghoul} and \textit{ghost} dominate the most common subject area of approximately 18 \% boys' programs (Topic~4).
 Although these might also be part of a fantasy world, both the bat, a bloodsucker that lives in the night, and a ghoul, i.e., a demon-like creature, commonly have rather negative and sinister connotations. The settings, \textit{cave} and \textit{woods}, also indicate a rather gloomy world. 
The example project in Figure \ref{fig:rq1topicexample} (ID55, Arena of Death) is assigned to Topic 4 by 97 \%, and was created by a boy. Sprite 1 (the default sprite) and the black bats remain static in the middle of the arena while the other bats and the ghouls move towards it. The script is the same for Bat1--3 and Ghoul1--3, and because of the missing termination condition the program does not terminate. There is no user interaction.
For girls, on the other hand, Topic 4 is one of the least popular topics occuring in only 4.67 \% of all girls' projects.
These large gender differences can be observed for several topics: While girls' projects often focus on dance and music (e.g., Topic 0, 15.02 \%), boys' projects seem to be more about games or football according to the terminology \textit{ball}, \textit{soccer} and \textit{goal} (Topic 6, 16.43~\%).

The distribution of the respective most popular topics of both genders thus already reveals some gender-stereotypical patterns established in society and a different interest in topics in \Scratch programs. 
These stereotypes seem to manifest themselves already in childhood \cite{palma2001women}.
The differences found in the topics reinforce this impression and indicate an interrelation that is also evident in \Scratch.
In all these cases the contrast between genders is striking, and clearly visible in the visualization in Figure~\ref{fig:lda2D}.

\subsubsection{Thematic Similarities}
There also exist thematic overlaps between girls and boys: Topics 2 and 3 are  similarly weighted between genders, though both topics are represented by only few projects that lack coherence in terms of content. Topic 8, which is overall the most common topic, and second and third most common for girls and boys (f: 16.37 \%, m: 14.38 \%), is about a circus with a ringmaster (``\textit{zirkusdirektor}'') and magician (``\textit{zauberer}''), various characters (\textit{girl} and \textit{boy}) and animals (\textit{monkey}, \textit{elephant}). This popularity is likely due to the introductory projects presented in the course, which suggests that girls and boys are either equally interested in the topic, or they were simply less creative for these projects. 

A similar observation applies to Topic 7, the overall fifth most popular topic for both genders (f: 9.35 \%, m: 11.64 \%): By means of a balancing tightrope walker (``\textit{seilt\"anzerin}'') as well as a cat pushing a \textit{ball} around, the children should get to know the \textit{motion} instructions better. In addition, the terms \textit{baseball} and \textit{basektball} opens up the subject area of ball sports, which could indicate children's interests in their leisure time. 
This topic seems to equally attract girls and boys through its diversity with the affinity for female tightrope walkers and stereotypically male ball sports.

Topic 0 is also a popular topic for both genders and the second most frequently mentioned topic in total. It covers the word fields around music (\textit{bass}, \textit{drum}, \textit{snare}) and celebration (\textit{dance}, \textit{celebrate}).

\summary{}{The most popular topics reflect gender stereotypical interests: girls' programs center on unicorns, celebrating, dancing, and music while boys prefer comparatively gloomy fantasy worlds with bats and ghouls as well as soccer. Topics involving circuses, birthday parties and animals appeal to both genders.}
\subsection{RQ2: Code Analysis}
\label{sec:rq2}

\begin{figure}
	\centering
	\includegraphics[width=\columnwidth]{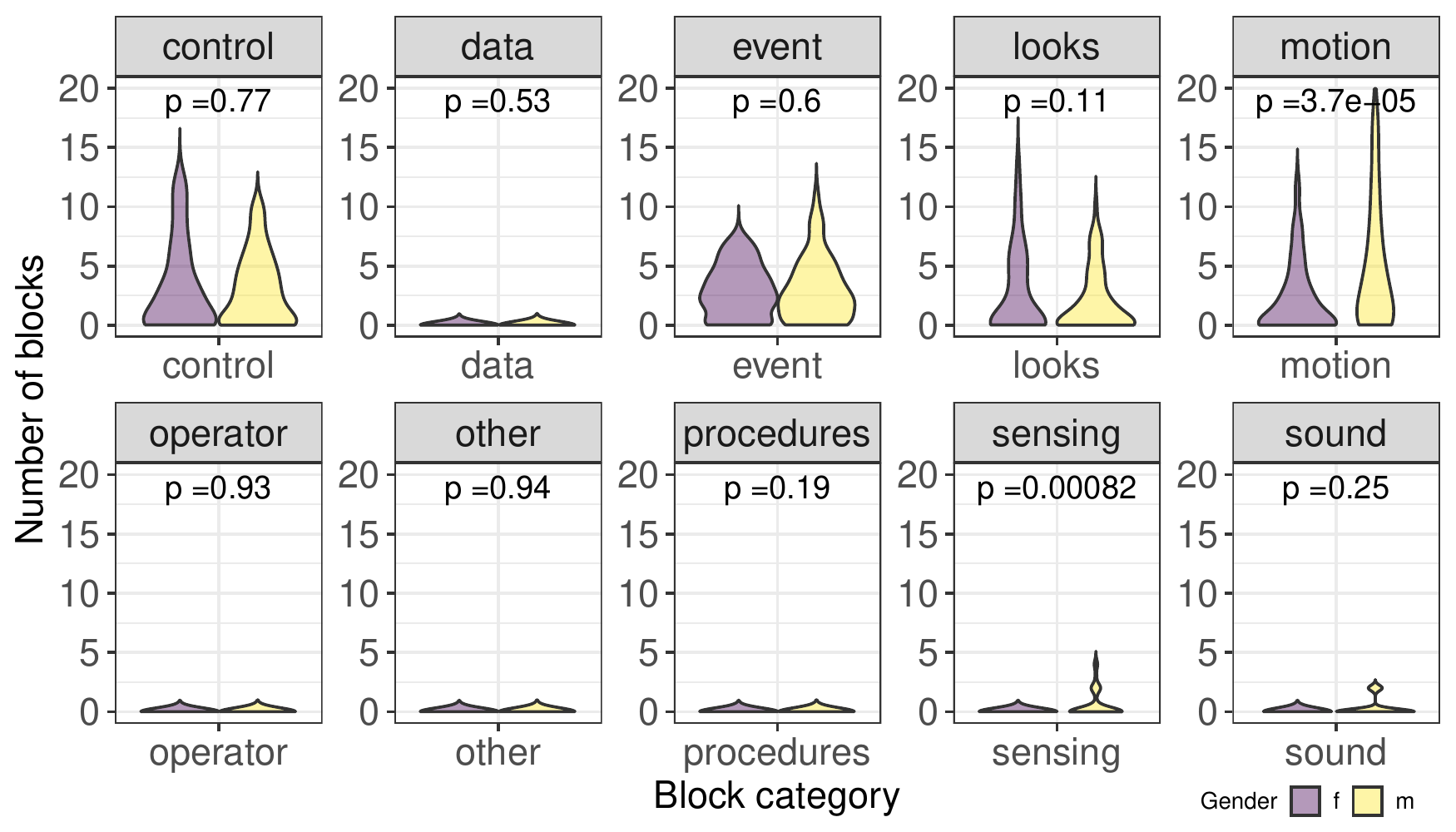}
	\vspace{-1em}
	\caption{The average usage of general block types per \Scratch program.\protect\footnotemark{}  The different uses of \emph{motion}  and \emph{sensing} blocks are statistically significant at $\alpha = 0.05$.}	
	\label{fig:generalOpcodes}
\end{figure}

 \footnotetext{Violins exclude outliers (i.e., 1.5 $\times$ interquartile range above the third quartile or below the first quartile) for readability; $p$-values calculated using Wilcoxon Rank Sum Test on the full data.} 
\begin{figure}
	\centering
	\includegraphics[width=\columnwidth]{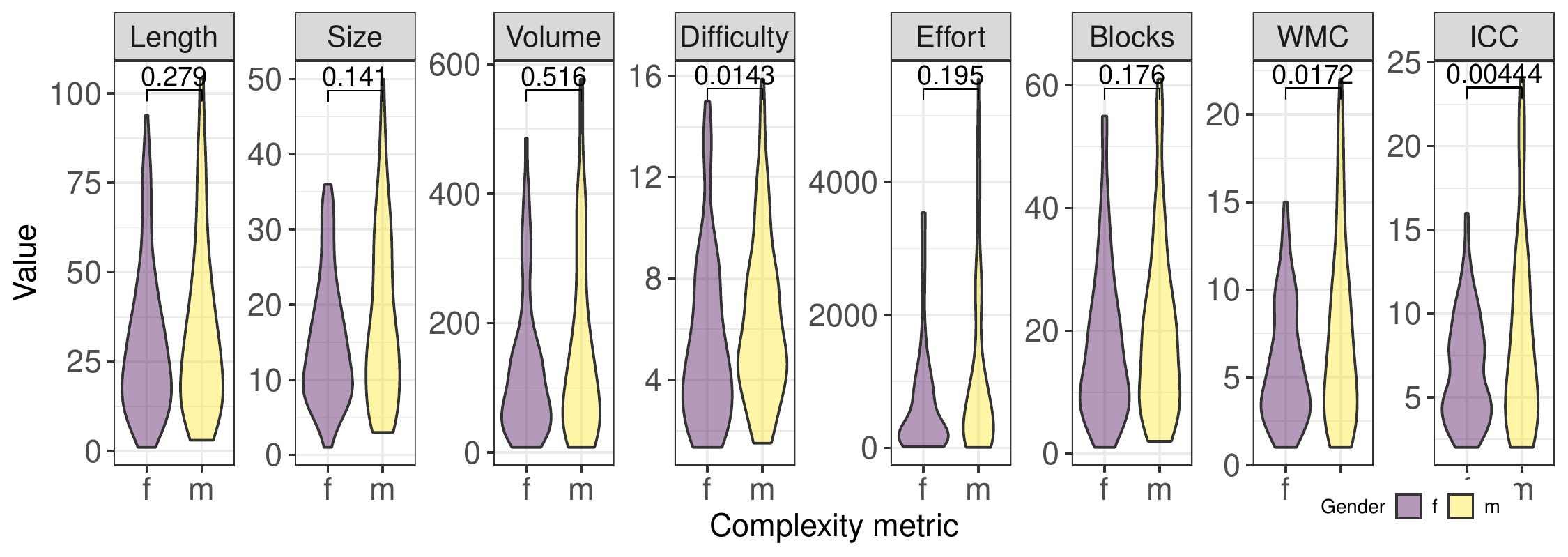}
	\vspace{-1em}
	\caption{The overall program complexity is measured by the Halstead metrics and the ICC.\protect\footnotemark{}  The Halstead difficulty, WMC and ICC are statistically significant at $\alpha = 0.05$.}
	\label{fig:codecomplexity}
\end{figure}

\footnotetext{See footnote 3.}

\begin{table}[t]
    \centering
    \caption{Top ten \Scratch block types per gender.}
    \label{tab:opcodesDetails}
    \resizebox{\columnwidth}{!}{%
    \begin{tabular}{rllrlllr}
    \toprule
    Rank  & Type & Block  & \# (f) &    Type & Block & \# (m) \\ \midrule
    1 & event& \begin{scratch} \blockevent{when \greenflag clicked} \end{scratch} & 565    
       & motion & \begin{scratch} \blockmove{move \ovalnum{} steps} \end{scratch} & 522  \\
    2 & control & \begin{scratch}\blockcontrol{wait } \end{scratch}& 502  
       & event &  \begin{scratch} \blockevent{when \greenflag clicked} \end{scratch} & 422 \\
    3 & looks & \begin{scratch}\blocklook{say \ovalnum{} for \ovalnum{} seconds}\end{scratch} & 434 
       & event & \begin{scratch} \blockevent{when \selectmenu{} key pressed} \end{scratch} & 337  \\
    4 & motion & \begin{scratch} \blockmove{move \ovalnum{} steps} \end{scratch}  & 243     
    & control & \begin{scratch} \blockinfloop{forever}{ \blockspace[0.4]} \end{scratch} & 316  \\
    5 & control & \begin{scratch} \blockinfloop{forever}{ \blockspace[0.4]} \end{scratch} & 174 
    & control & \begin{scratch}\blockcontrol{wait } \end{scratch}& 233  \\
    6 & control &\begin{scratch}\blockrepeat{repeat \ovalnum{}} {\blockspace[0.4]} \end{scratch} & 148   
     & looks & \begin{scratch}\blocklook{say \ovalnum{} for \ovalnum{} seconds}\end{scratch} & 174  \\
    7 & sound & \begin{scratch}  \blocksound{\selectmenu{}}  \end{scratch}& 147   
    & control  & \begin{scratch} \blockif{if \boolempty[2em], then} {\blockspace[0.4]} \end{scratch} & 168 \\
    8 & looks & \begin{scratch} \blocklook{costume \selectmenu{} } \end{scratch} & 113    
    & motion & \begin{scratch} \blockmove{point in direction \ovalnum{} } \end{scratch}  & 159  \\
    9 & looks &\begin{scratch} \blocklook{switch costume to \selectmenu{} } \end{scratch}& 113   
    & control &  \begin{scratch}\blockrepeat{repeat \ovalnum{}} {\blockspace[0.4]} \end{scratch} & 144 \\
    10 & motion &\begin{scratch} \blockmove{turn \turnright{} \ovalnum{} degrees} \end{scratch} & 104   
    & motion & \begin{scratch}\blockmove{turn \turnright{} \ovalnum{} degrees} \end{scratch} & 123  \\ \bottomrule
    \end{tabular}%
}	
    \end{table}

\begin{table}[tb]
\renewcommand{\arraystretch}{0.7}
\small
\centering
\caption{Usage of programming concepts and their corresponding block type in \Scratch programs.}
\label{tab:programmingconcepts}
\begin{tabularx}{\linewidth}{@{}llXrr@{}}
\toprule
Concept & Type & Block & \# (f) &  \# (m) \\ \midrule
\multirow[t]{4}{*}{Conditional} & control & if & 77 & 168 \\ 
 & motion  & if on edge bounce & 54 & 91 \\
 & control & if else & 3 & 8 \\
 & total & &134 & 267 \\ \midrule
\multirow[t]{3}{*}{Coordination} & event & broadcast & 4 & 10 \\ 
 & event & when broadcast received & 4 & 17 \\ 
 & event & broadcast and wait & 4 & 17 \\
 & control & wait & 502 & 233 \\ 
 & control & wait until & 0  & 0 \\ 
 & total & &510 & 260 \\ \midrule
\multirow[t]{3}{*}{Iteration} & control & repeat & 147 & 144 \\ 
 & control & repeat until & 0 & 1 \\
 & control & forever & 174 & 316 \\ 
 & total & & 321 & 461 \\ \midrule
\multirow[t]{3}{*}{Variables} & data & change variable by & 4 & 23 \\ 
 & data & set variable to & 18 & 5 \\  
 & data & variable & 0 & 4 \\ 
 & data & show variable & 0 & 0 \\ 
 & data & hide variable & 0 & 0 \\
 & total & & 22 & 32 \\ \bottomrule
\end{tabularx}
\end{table}

\begin{table}[tb]
    \centering
    \renewcommand{\arraystretch}{0.8}
    \caption{The ten most common code smells in \Scratch programs of both genders, normalized per project.}
    \label{tab:smells}
    \begin{tabularx}{\linewidth}{@{}rlrXr@{}}
    \toprule
    Rank & Code Smell (f) & \# (f) & Code Smell (m) & \# (m) \\ \midrule
    1 & Duplicate Sprite & 10.22 & Duplicate Sprite & 5.15 \\
    2 & Empty Sprite & 2.38 & Missing Init. & 3.52 \\
    3 & Missing Init. & 1.66 & Sprite Naming & 1.76 \\
    4 & Sprite Naming & 1.63 & Stuttering Movement & 1.31 \\
    5 & Clone Type 3 & 0.71 &  Empty Sprite & 1.22 \\
    6 & Clone Type 2 & 0.40 & Clone Type 1 & 0.89 \\
    7 & Dead Code & 0.23 & Clone Type 3 & 0.80 \\
    8 & Empty Script & 0.19 & Dead Code & 0.43 \\
    9 & Long Script & 0.16 & Clone Type 2 & 0.41 \\
    10 & Clone Type 1 & 0.14 & Miss. Pen Up / Erase All & 0.24 \\\bottomrule
    \end{tabularx}
    \end{table}

While the first research question looked at the overall topics of the programs implemented, in this research question we now turn to the insides of the programs and look at the source code. 
Figure~\ref{fig:generalOpcodes} summarizes the distribution over different categories of the blocks used, and Table~\ref{tab:opcodesDetails} shows the top 10 blocks used for each gender.
Table~\ref{tab:programmingconcepts} summarizes which programming concepts are covered by these blocks.
Figure~\ref{fig:codecomplexity} summarizes different complexity metrics for the programs.
Table~\ref{tab:smells} shows the top 10 code smells for each gender.

%
%
%
%
%

\subsubsection{Code Differences}
\label{sec:rq2diffs}

According to Figure~\ref{fig:generalOpcodes}, the largest number of blocks in
the boys' programs (30\% of all blocks used) are from the \textit{motion} category.
Table~\ref{tab:opcodesDetails} confirms that the most frequent block is
\textit{move steps} which is responsible for moving sprites, and further popular
blocks are \textit{point in direction} and \textit{turn right}. While there is no
significant difference in the overall number of \textit{event} blocks used,
Table~\ref{tab:opcodesDetails} shows that boys use \textit{when key pressed}
frequently. Together
with the significantly higher number of \textit{sensing} blocks, and a slightly higher
number of \textit{event} blocks, this is indicative of interactive, game-like programs.

Girls use considerably more \textit{looks} blocks---26\% of all the blocks they use are
from this category. Table~\ref{tab:opcodesDetails} shows that, for girls, the
most frequent \textit{looks} block are \textit{say for secs}, \textit{costume}, and
\textit{switch costume to}, which cause sprites to produce dialog using speech
bubbles and change appearance. These blocks are essential components of
story-like projects, with dialogue between sprites in the program.
However, the use of \textit{sensing} blocks (which includes asking the
user questions) is done significantly less often by girls compared to boys. 

In the programs of the introductory course, 20 different types of blocks were introduced. There are also differences in the additional individual blocks of the children. There are 21 block types used only by girls and 20 block types used only by boys. The most common types of blocks that only girls used are: \textit{rest for beats}, \textit{change tempo}, \textit{reporter string number}. 
The most common block types that only boys have used are: \textit{change pen hue by}, \textit{change y by}, \textit{random}. 


Although there are no significant differences in the overall number of \textit{control} and \textit{event} blocks used, Table~\ref{tab:opcodesDetails} shows that the blocks within these categories do differ: Girls mainly use \textit{when flag clicked} event handlers, and simpler control structures (\textit{forever}, \textit{repeat}, \textit{wait}), whereas \textit{if} makes an appearance in the boys' top 10 blocks. Table~\ref{tab:programmingconcepts} confirms that conditional statements (\textit{if}, \textit{if on edge bounce}, \textit{if else}) are substantially more frequent in boys programs. This is likely another artifact of them preferring to implement games, where the gaming flow requires those concepts as well as dynamic input more often. 
Although the loop-statements used by boys and girls are similar, Figure~\ref{tab:programmingconcepts} shows that boys use loops more frequently, whereas girls seem to prefer programs with sequential flow.

These differences in the control structures imply differences in the complexity of the resulting programs.
The Halstead metrics in Figure~\ref{fig:codecomplexity} all show slightly higher values for the boys' programs: There are no significant differences in terms of length (total number of words used) or size (total number of unique words), and thus also no significant difference in terms of the resulting Halstead volume. The Halstead effort, which is meant to estimate the amount of mental effort required to recreate the software, is substantially higher for boys' programs, and the Halstead difficulty is significantly higher for boys' programs (p=0.01).

The weighted method count (WMC) shows the sum of cyclomatic complexities of all scripts in a program, while the interprodedural cyclomatic complexity (ICC) is calculated on the interprocedural control flow graph. In both cases there is a significant difference observable (WMC: p=0.02, ICC: p < 0.01), suggesting that boys create more complex programs (Figure~\ref{fig:codecomplexity}). This is in line with the observation that boys use more if-statements (as shown in Table~\ref{tab:opcodesDetails}).

The concept of \Scratch eliminates syntax errors, but there are still many other types of errors and bugs \cite{hermans2016a, fradrich2020}. Since the children's programs are relatively small and less complex, fewer code smells are likely to occur. They do, however, corroborate our previous observations on the difference in the programming concepts applied: 
Girls' projects contain the smells \textit{Duplicate Sprite} and \textit{Empty Sprite} twice as often (Table~\ref{tab:smells}). This may be due to the fact that in animations and stories many sprites serve as decoration and are therefore empty. 
Boys' projects, on the other hand, contain the smell \textit{Missing Initialization} about twice as often. \textit{Missing Initialization} is mainly due to the category game, because the position has to be changed, but is never set. 
As Table~\ref{tab:smells} shows, the smell \textit{Stuttering Movement} in the programs of boys is striking, which points to programs with a lot of interaction of the user and thus to the project category game. The smell is also one of the most common bug patterns in \Scratch programs~\cite{fradrich2020}.

%
%
%

\subsubsection{Code Similarities}

Although there are differences in the types of blocks used, and the complexity of the structures that connect these blocks, the programs are comparable in size (p=0.24): Boys' programs contain on average almost 27 blocks, while girls' programs contain slightly less, around 22 blocks. 
While we saw a large difference in terms of Halstead effort, the size-related Halstead metrics in Figure~\ref{fig:codecomplexity} suggest that programs are comparable in terms of how diverse blocks are within the programs.

According to Figure~\ref{fig:generalOpcodes}, two of the largest categories of
blocks are related to \textit{control} (if-statements, loops), \textit{events} (event handler
blocks, with comparable distributions between girls and boys. Even though there
are differences between which specific blocks are used as discussed above, the
interleaving of control and non-control blocks shows a comparable desire to
control the program execution.

Blocks of categories other than control, event, looks, and motion are used less
frequently by both, girls and boys; in particular we see almost no usage of
blocks from the data, operator, and procedures categories, which cover concepts
that were not covered in the introductory teaching material.

Compared with the different block types used in the sample projects from the introductory course, the ratio of blocks differing from this in the programs of girls and boys is almost identical (f: 28.57 \%, m: 27.24 \%) as girls add another 59 block types to their programs and boys add 60 more. Both genders are equally skilled at exploring unfamiliar block types on their own as they extended their programs proportionally equally, but conceptually differently as described in Section \ref{sec:rq2diffs}. 
As Table~\ref{tab:smells} shows, we also observe some common code smells that are independent of gender:
The smell \textit{Sprite Naming} means that the default name of the sprite is used or that the same name is only used in an incremented form (e.g., cat1, cat2). Thus, girls and boys invented their own names for their sprites with similar frequency.


\summary{}{The usage of code blocks in \Scratch programs of girls and boys differ. Girls use blocks that are suitable for stories and animations, such as \textit{looks}, and boys mainly use game-specific blocks such as \textit{motion}. Boys apply more programming concepts such as conditional statements or iterations. The type of code smells differ only slightly as the code smells are mainly due to the specifics of the project type.}

\section{Discussion}
\label{sec:discussion}

\subsection{Interrelation of Topics and Code}

\begin{figure}[tb]
	\centering
	\includegraphics[width=\linewidth]{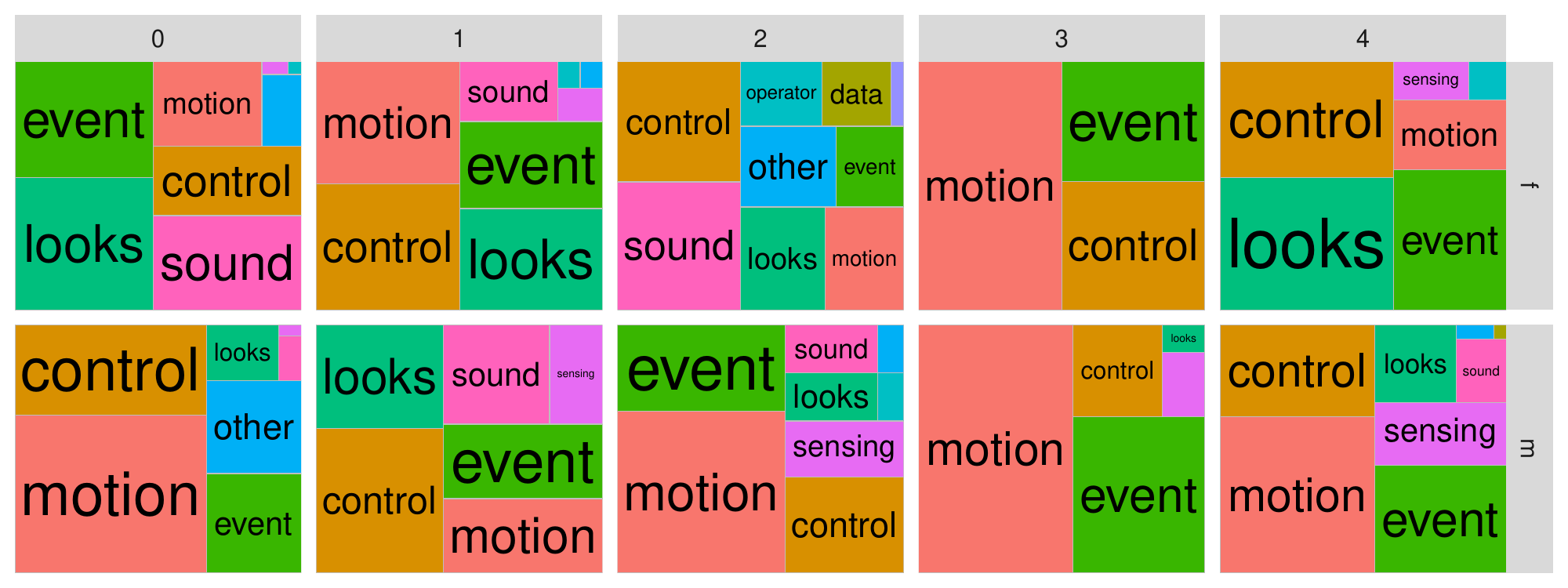}
	\includegraphics[width=\linewidth]{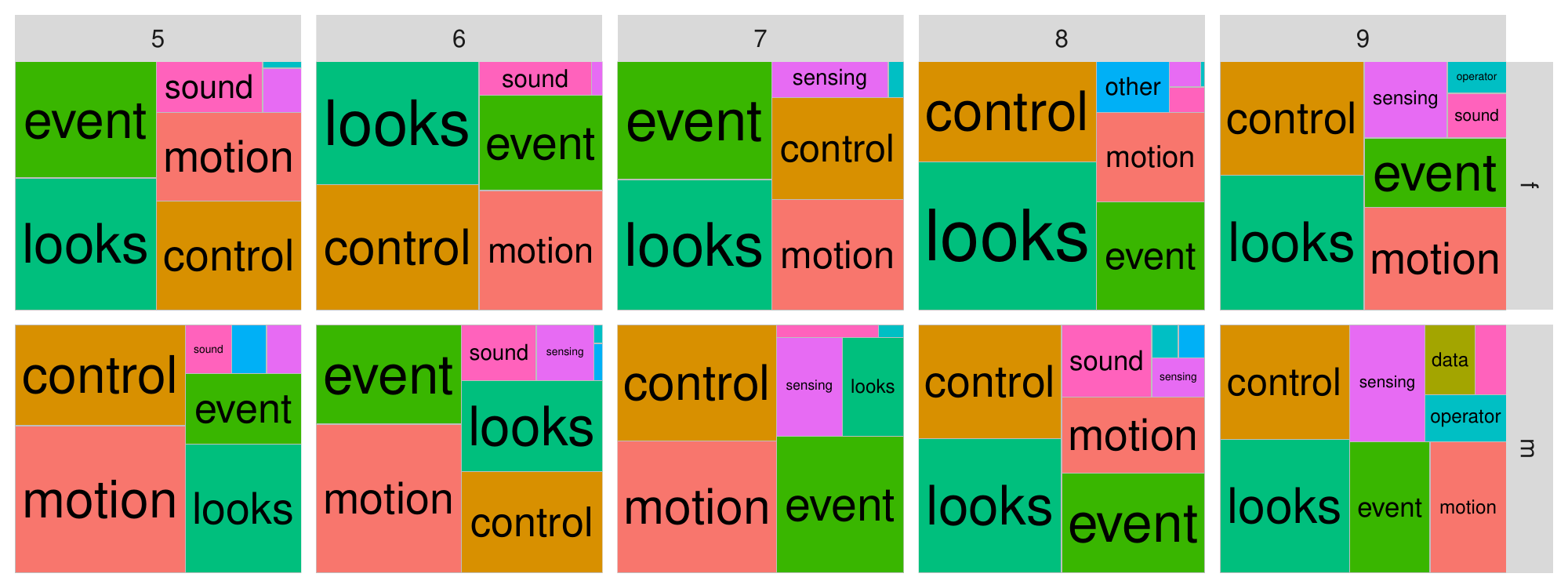}
	\caption{Distribution of block types among topics/genders.}
	\label{fig:topictypes}
\end{figure}

\begin{figure}[tb]
	\centering
	\includegraphics[width=\linewidth]{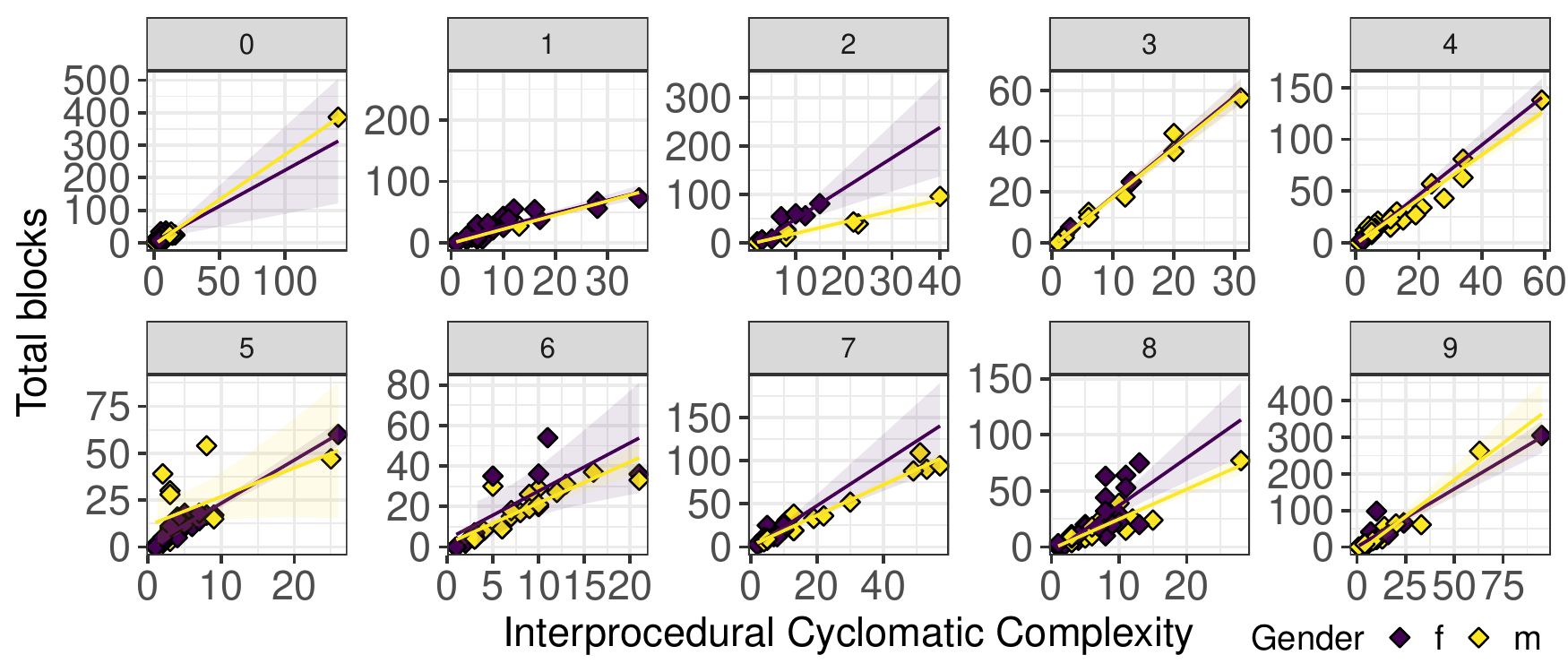}
	\caption{For each topic of the LDA model, their programs are shown by their complexity in relation to their size.}
	\label{fig:topiccomplexity}
\end{figure}

We observed differences in the topics that girls and boys use when creating
\Scratch programs (RQ1) and we observed that girls are less likely than boys to
use programming concepts such as iterations and conditions (RQ2). Ideally, we want girls to be able to implement topics they like, but produce code that covers programming concepts to the same extent as boys' programs. Figure
\ref{fig:topictypes} therefore shows the distribution of block types among the
topics, and Figure \ref{fig:topiccomplexity} shows the relation of complexity and size of the projects of each topic. This allows us to investigate whether we observe any differences between the code produced for different topics.

Considering the distribution of block types for girls in Figure~\ref{fig:topictypes}, we see that the use of \textit{motion} blocks is similarly sparse across all topics except Topic 3, which in fact is a strong outlier for girls and boys (f: 50 \%, m: 53 \% of all blocks used in this topic are \textit{motion} blocks). However, this does not imply the topic is well suited to guide girls to produce programs with more interactive behavior: As Figure \ref{fig:topiccomplexity} demonstrates, these projects contain almost no code, but instead consist of modified or self-painted backgrounds. For example, one girl's program (ID1) contains a cat moving towards a ball. This observation is not gender-specific, as the boys' projects for this topic also contain very few blocks. For example, a boy's project (ID120) contains several lions and cats moving back and forth. Thus, even though boys and girls use the topic in a similar way, it does not appear to foster exercising programming concepts.

\begin{figure}[tb]
\centering
\subfloat[\label{girlreverse}This girls' program (ID126) is 94 \% assigned to Topic 4.]{\includegraphics[width=0.46\linewidth]{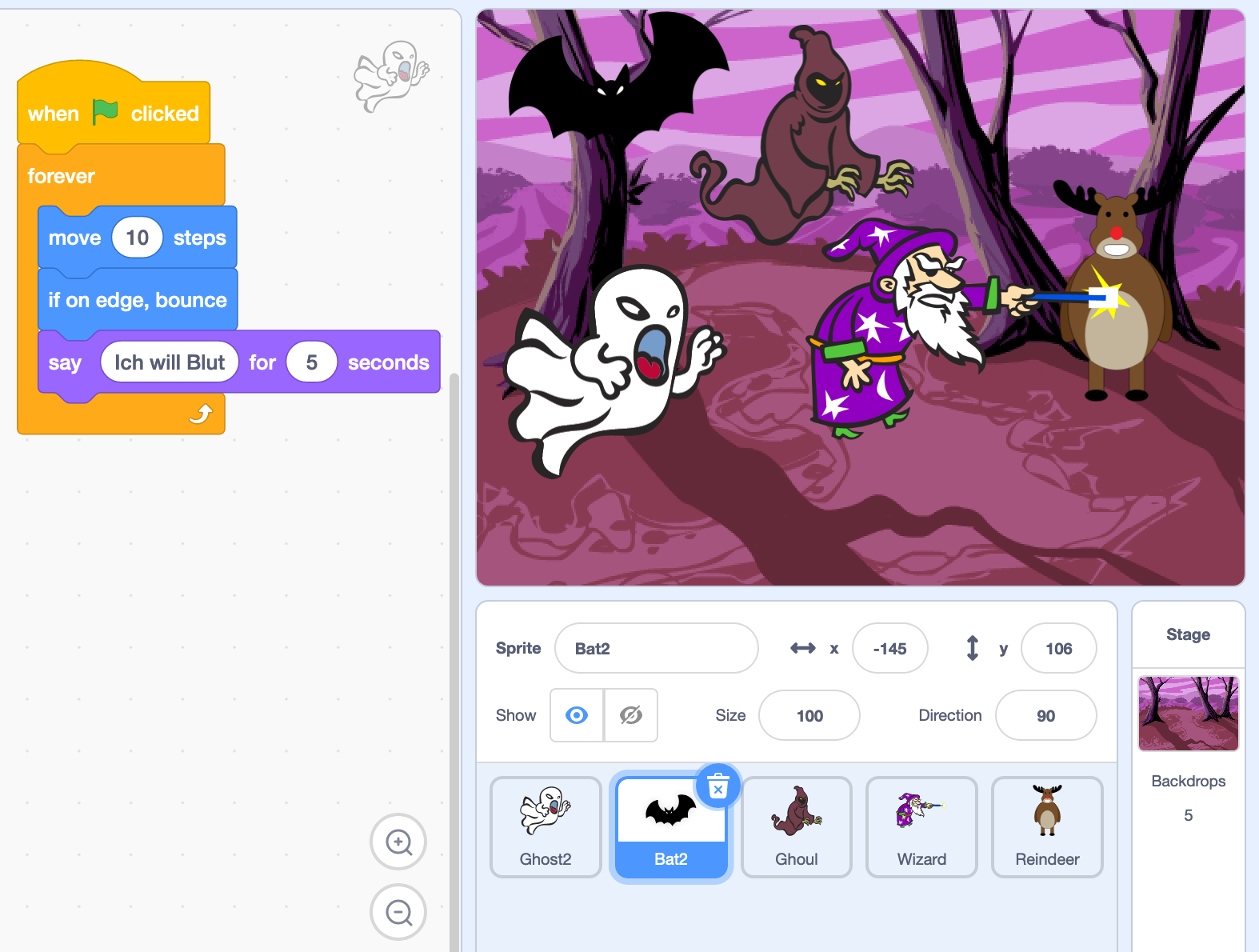}}
\hfill
\subfloat[\label{boyreverse}This boys' program (ID100) is 59 \% assigned to Topic 1.]{\includegraphics[width=0.46\linewidth]{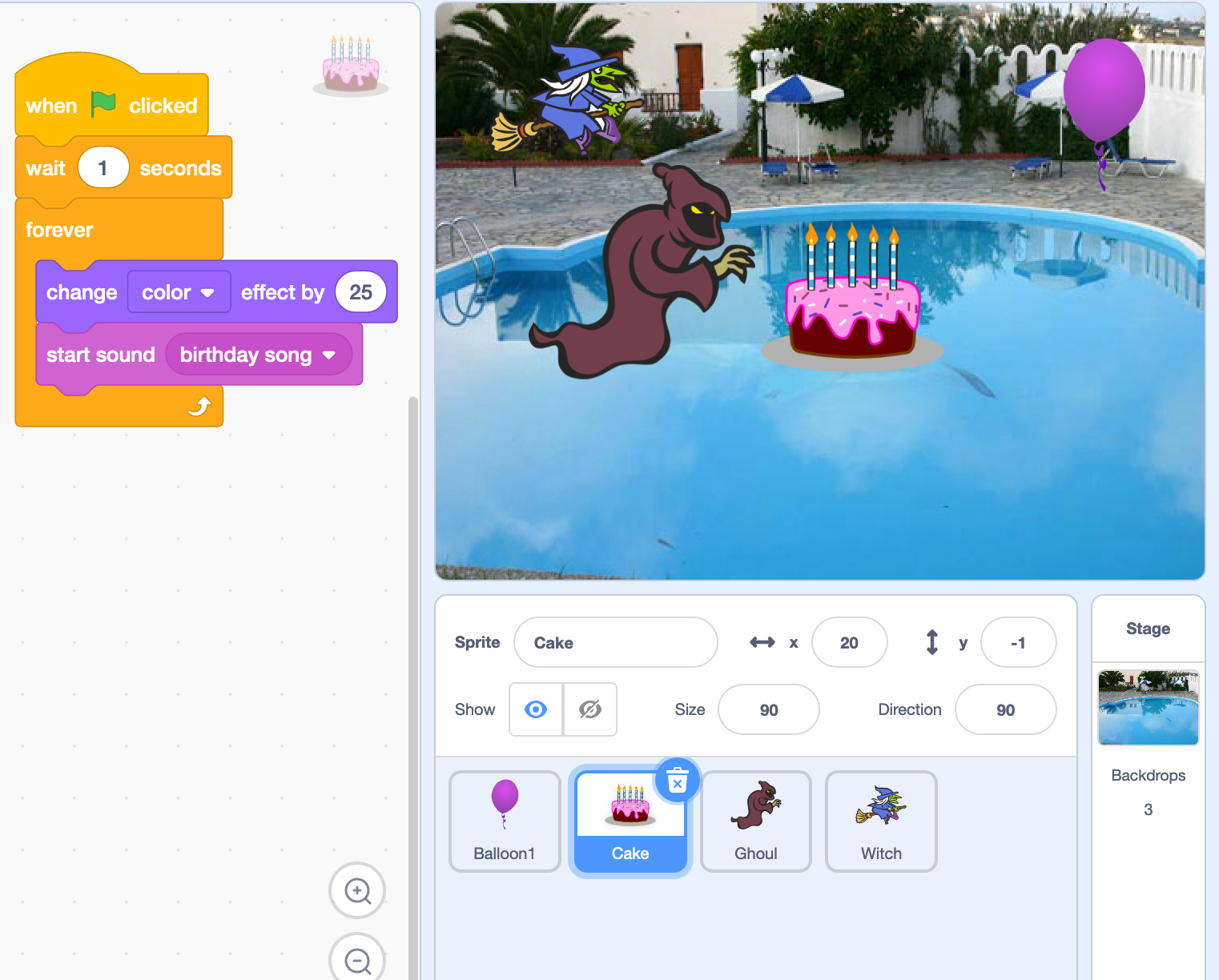}}
\hfill
\subfloat[\label{circus1} This girls' program (ID26) is 98 \% assigned to Topic 8.]{\includegraphics[width=0.46\linewidth]{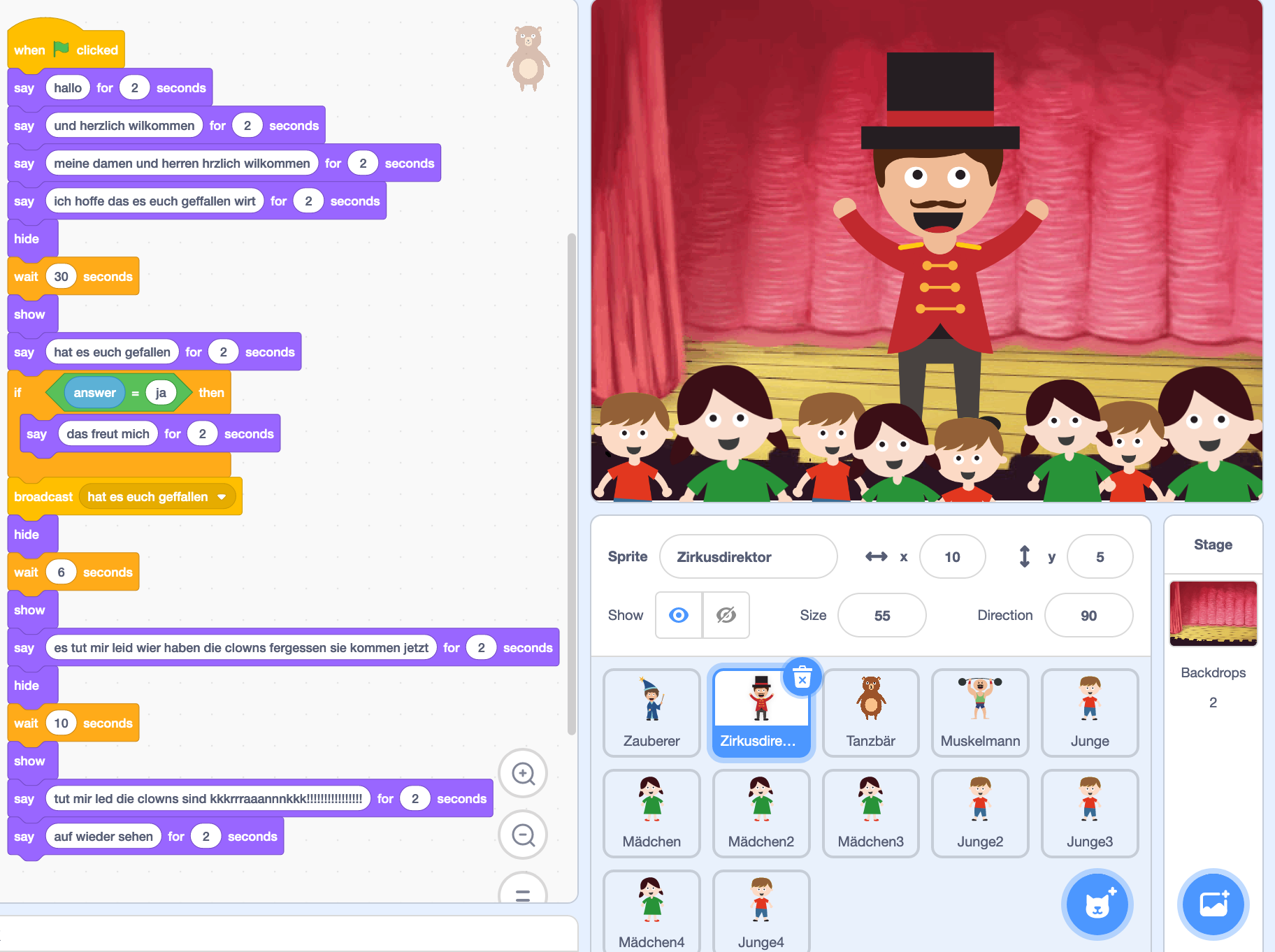}}
\hfill
\subfloat[\label{circus2} This boys' program (ID19) is 96 \% assigned to Topic 8.]{\includegraphics[width=0.46\linewidth]{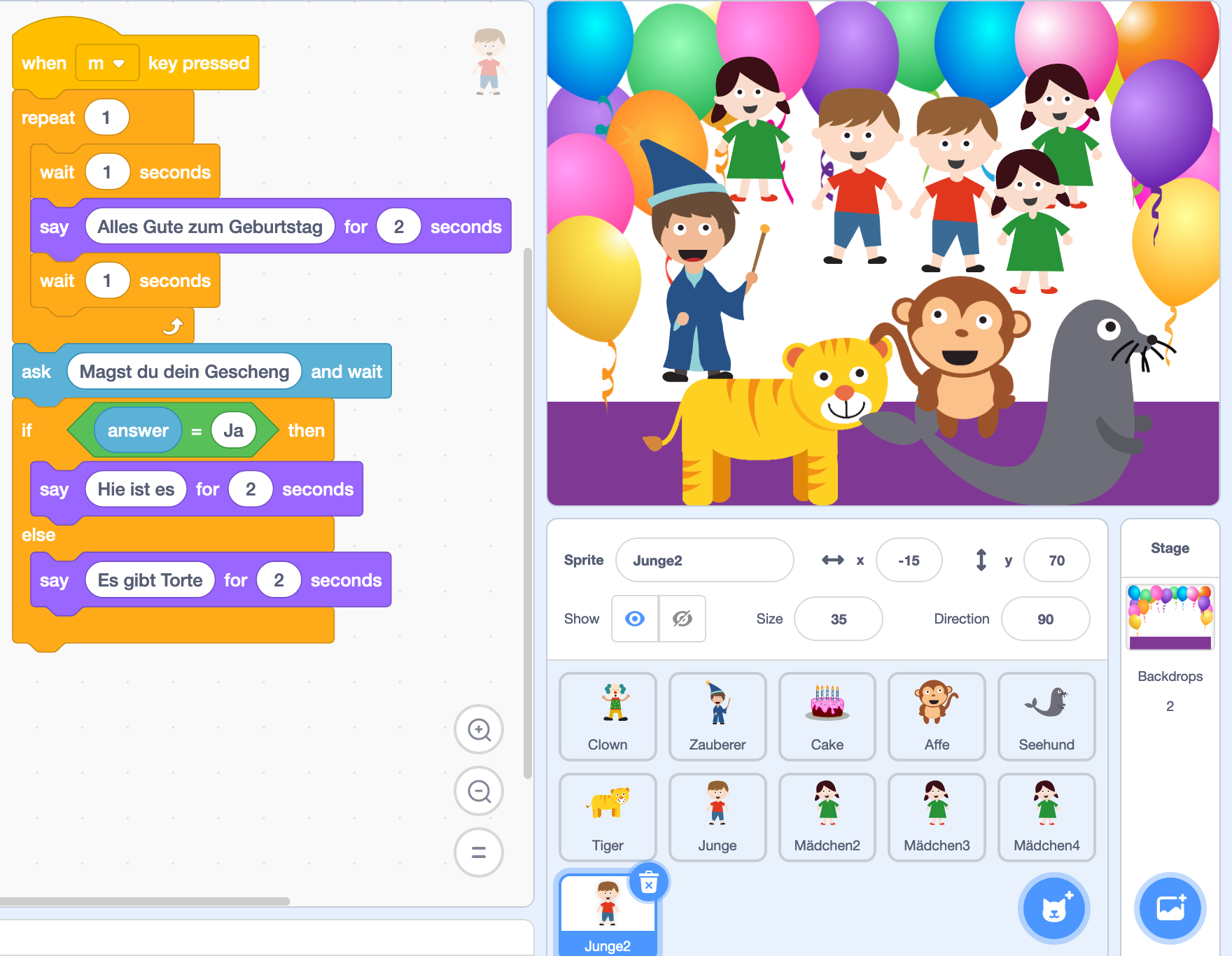}}
\caption{One representative example for each of the top-topics of boys (\ref{girlreverse}) and girls (\ref{boyreverse}) as well as two representative examples of the universally popular circus topic (\ref{circus1}, \ref{circus2}).}%
\label{fig:examplesreverse}%
\end{figure}

To see whether girls are more likely to produce game-like programs for topics boys prefer, we examine Topic 4, the most popular topic among boys (Fig. \ref{fig:topictypes}). Boys' projects in this topic are primarily games that require many \textit{motion} and \textit{control} blocks (f: 25 \%, m: 22 \% of all blocks used in this topic). These are games whose presentation and mechanics are reminiscent of classic maze arcade games like Pac-Man, Ghost Hunt or platform games. Figure \ref{fig:topiccomplexity} shows that girls' projects in this topic are mostly small and simple (average blocks: 11, ICC: 5.37), while boys projects are both larger and more complex (average blocks: 28, ICC: 13.51). The dominant category of blocks used by girls for this topic are \textit{looks} blocks, which account for 32 \% of all blocks in this topic (Fig. \ref{fig:topictypes}). For example, Figure \ref{girlreverse} shows an example girl's project for this topic: Even though it contains a Bat sprite, this sprite just says ``I want blood'' while the other sprites are empty, like in many of the girls' programs. Thus, even for popular boys' topics, the girls' projects follow the pattern of animation programs without interactions. According to the use of blocks such as \textit{looks} and \textit{sound} as shown in RQ 2, it indicates that girls' projects are predominantly of the animation project type.

We do, however, observe in Figure \ref{fig:topictypes} some differences in the projects that boys produce for the topics dominated by girls, compared to other topics: For both genders the majority of programs in the topics popular among girls (Topic 0, 1, 5) fall into the animation or story type, which requires little to no user interaction. Figure \ref{fig:topiccomplexity} also shows that within these topics, boys’ projects are similar in size and complexity to girls’ projects, while in most other categories, boys' projects are more complex. 
In Topic 1 (Fig. \ref{fig:topictypes}), the most popular girl topic, the boys' projects also use a relatively large number of \textit{looks} blocks (18\% in Topic 1 versus 8\% in Topic 4 of all blocks used).
The boy’s program with the highest assignment to Topic 1 (Fig. \ref{boyreverse}) seems to be about a birthday party without interaction. However, there are only three boys' programs in this topic.
%


Topic 8 occurs with similar frequency for girls and boys and is overall the most common topic, as it represents the circus topic used in the instructional programs. For this topic, the distribution of block types seems relatively balanced across genders (Fig. \ref{fig:topictypes}). It has the highest proportion of \textit{looks} blocks across all topics for girls and boys (f: 37 \%, m: 27 \% of all blocks used in this topic) and a fair amount of \textit{control} elements (f: 25 \%, m: 22 \% of all blocks used in this topic). Fig. \ref{fig:topiccomplexity} shows that the projects are comparable in terms of size (approximately 21 blocks used for both genders), although the boys' projects again tend to be slightly more complex overall (ICC per project f: 6.46, m: 8.71).
Considering sample projects (Fig. \ref{circus1}, \ref{circus2}), the narrative as well as the code used is similar: there is a stage with many sprites (sprites per project f: 4.5, m: 5.0) that interact with each other through communication, where wait and if blocks are used, in order to be able to react to certain input from the user, for example whether the performance was liked. 
This suggests that using the topic featured in the initial work materials may support a more balanced use of programming concepts.

The distribution in Figure \ref{fig:topiccomplexity} shows that outliers exist for all topics and girls and boys. These extreme outliers are usually the result of duplicated sprites, which is a common practice for both genders (Table~\ref{tab:smells}); for example, two most extreme projects are a boy's program (ID 49) with almost 368 blocks in Topic 0 (70 \%) and a girl's program (ID125) with 305 blocks in Topic 9 (98 \%), both of which contain dozens of duplicated sprites.

\subsection{Recommendations}

\subsubsection{Suggestions for Code Elements} 

It was shown that girls prefer the project types animation and story, regardless of the topic, and thereby implement fewer programming concepts (Table \ref{tab:programmingconcepts}), which leads to less complex programs (Fig. \ref{fig:codecomplexity}). There is a risk that girls who implement the seemingly simpler concepts will remain at a somewhat lower level of knowledge and skills, which can be detrimental to their enthusiasm and the unfolding of their full potential.
To address this, programming tasks should be designed that provide sufficient room for all children's thematic preferences and the application of programming concepts. The aim should be to design challenging tasks but also to further increase the girls' motivation by using elements of their favoured project types.

One way could be to foster the use of cross-category elements in programming tasks, for example using blocks that particularly appeal to girls in games. This could be realized, for example, by having a sprite change its costume, say something or make a sound on certain events in the game as \textit{switch costume to} and \textit{sounds menu} are two of the most popular blocks of the girls (Table \ref{tab:opcodesDetails}). 

Another way to lead girls to using more complex programming concepts might be to provide specifications in open programming scenarios, such as \textit{``the main character in your program should be controlled by the user''}, which implies reacting to user input and thus using if-conditions or multiple events. In this case, the topic of the program can be left completely open, or a rough thematic framework can be given, e.g., a visit to the zoo or a day at the beach. 

An alternative could be to develop tasks based on short stories and in this way evoke the desired programming concepts. The storyline could suggest certain programming concepts, e.g., \textit{``no matter where the mermaid moved, the fish followed her''} and still be appealing to girls. Free spaces could be left in the stories for them to fill in creatively, such as \textit{``the crab was very happy to meet the octopus''}. This phrase could be implemented in very different ways---the crab could say something joyful, jump in joy etc.

  
The starter projects on the \Scratch website\footnote{https://scratch.mit.edu/starter-projects} also show that certain topics are more suitable for certain project types. For example, dance and music topics, such as the Dance Party project, are offered as animation, while the games may be less appealing to girls. Here, one possibility would be to address girl-specific topics or at least more neutral topics to motivate girls to implement games as well. 

\subsubsection{Topics as Motivators}
The topics identified from RQ1 exemplify that the areas of interest of girls and boys differ in the implementation of \Scratch projects, but that there are also thematic overlaps. The most popular topic in boys' programs is particularly interesting as it relates to the introductory \Scratch program \textit{Ghostbusters} in the first module of Code Club\footnote{https://projects.raspberrypi.org/en/projects/ghostbusters} --a large initiative to get children interested in programming. 
One could eventually establish a connection here: Boys in particular are addressed with this content and also may have been inspired by this task. Girls may have less desire to implement the programming task, simply because they are less interested in the topics compared to others.

Therefore, a motivator would be to thematically adapt an introductory task to girls' interests: Besides \textit{Ghostbusters}, an example program with more ``girly'' topics containing \textit{Unicorns} or an equivalent set of topics from pop culture, analogous to \textit{Ghostbusters} should be offered as a springboard~\cite{corneliussen2016kids}.
When children are working on one project in a session, providing a gender-neutral topic is beneficial.  
This could also have a positive effect on attracting interest in more diverse groups, especially for events such as introductory programming courses or \textit{Take Our Daughters and Sons to Work Day}. In this context, it is important to ensure that no gender stereotypes are reproduced in these introductory programs, but that universal topics provide an entry point for further enthusiasm in programming~\cite{rubegni2020}.

Just as with the introductory tasks at Code Club, it is important to pay attention to thematic diversity in learning environments. 
A completely free design of their programs would be most beneficial for children~\cite{roque2016}. If this is not possible due to certain learning goals or if programs are presented as examples, care must be taken to ensure that the content of the programs can potentially inspire all target groups, and thus arouse interest in programming and CS. 

\section{Conclusions and Future Work}
\label{conclusions}

In this paper, we identified gender differences and similarities in 317 \Scratch programs using automated topic and code analysis. 
A topic analysis revealed that girls' programs revolve around unicorns, music and dance, while boys prefer more gloomy fantasy worlds with bats and ghouls as well as soccer or basketball. 
%
While both genders use similar numbers of blocks to implement these programs, usage varies: Girls mainly use \textit{looks} and \textit{sound} blocks, which indicate the category of story and animation, while boys predominantly use \textit{motion} blocks, which are game-specific elements. Boys are also much more likely to use essential programming concepts such as conditional statements or iterations, resulting in slightly higher complexity. 
These findings suggest a better internalization of programming concepts by boys.
%
%
These results suggest that teaching materials 
must be adapted in such a way that all genders are equally challenged and motivated~\cite{p.rose2020}. It is also important to address the special needs and interests of girls in order to promote them in particular. This also has an impact on the design of starter projects on the \Scratch website and the programming examples in the Code Club.
%
%
Furthermore, it is important to create an awareness for the interests of the users and the accompanying stereotypes among the educators to ensure that these are not reproduced in the teaching concepts~\cite{rubegni2020}. 
For future work it is relevant to consider all facets of diversity and to expand the analyses to include not only gender, but also ethnicity, social background, or disabilities~\cite{richard2016, roque2016}. A deeper understanding of behavior in \Scratch programming~\cite{fields2017} will enable an effective learning environment for all children equally, which helps to strengthen their computing-related self-concept and thus their longer-term interest in CS.

\vspace{-0.5em}
\begin{acks}
\vspace{-0.3em}We thank Alexandra Funke for her contribution to the data collection. This work is supported by the Federal Ministry of Education and Research
through project ``primary::programming'' (01JA2021) as
part of the ``Qualitätsoffensive Lehrerbildung'', a joint initiative of the
Federal Government and the Länder. The authors are responsible for the content
of this publication.
\vspace{-0.5em}
\end{acks}

 \bibliographystyle{ACM-Reference-Format}
 \bibliography{references}

\end{document}